\documentclass[a4paper]{article}
\usepackage{graphicx}
\usepackage{amsmath}
\usepackage{amsfonts}
\usepackage{amssymb}

\begin{document}

\title{Maximum pull out force on DNA hybrids}
\author{P.\ G.\ de Gennes\\Coll\`{e}ge de France, 11 place Marcelin Berthelot\\75231 Paris Cedex 05, France\\E.mail: pgg@espci.fr}
\maketitle
\begin{abstract}
We discuss theoretically the force $F$ between two colloidal particles, each
of them carrying one single strand DNA. The two strands are complementary
\ only on a finite sequence of ($\ell$) consecutive base pairs. We define an
adjustment length $\varkappa^{-1}$ (a few base pairs): in the adjustment
regions near both ends of the paired region, the tension is still mainly on
one single strand.\ But in the central part (for $\ell>\varkappa^{-1}$) the
two backbones are equally loaded. This leads to a rupture \ force $F_{c}$
\ increasing \ linearly with $\ell$ for $\ell\varkappa<1$, and saturating for
$\ell\varkappa>1.$
\end{abstract}

\bigskip

\qquad\qquad\qquad\qquad\qquad\qquad\textbf{Abstract fran\c{c}ais}

\bigskip

\qquad\qquad{\large Force maximum de tirage sur un ADN\ hybride}

\bigskip

Deux particules collo\"{i}dales, chacune portant une cha\^{i}ne greff\'{e}e
d'ADN, et les deux cha\^{i}nes pouvant s'hybrider, peuvent se ponter. On
discute th\'{e}oriquement la r\'{e}sistance en tension de ce pontage, lorsque
l'hybrydation porte sur une s\'{e}quence de $\ell$ bases cons\'{e}cutives. Il
appara\^{i}t une longueur d'ajustement $\varkappa^{-1}$ (de l'ordre de
quelques paires de base).\ Dans une r\'{e}gion d'ajustement (de taille
$\varkappa^{-1}$) pr\`{e}s de chaque extr\'{e}mit\'{e}, la structure est
distordue. Par contre, dans la r\'{e}gion centrale (pour $\varkappa\ell>>1$)
les deux cha\^{i}nes supportent des tensions \'{e}gales, et la structure n'est
pas perturb\'{e}e. Ceci conduit \`{a} une force de rupture $F_{m}$ qui
augmente lin\'{e}airement avec $\ell$ pour $\varkappa\ell<1$, et qui sature
pour $\varkappa\ell>1.$

\bigskip

\qquad\qquad\qquad\qquad\qquad\qquad\textbf{Texte court en fran\c{c}ais}

\bigskip

Divers syst\`{e}mes de reconnaissance des acides nucl\'{e}iques sont
fond\'{e}s sur des cha\^{i}nes \`{a} un brin,\ greff\'{e}es sur une surface,
et oppos\'{e}es \`{a} des cha\^{i}nes partiellement compl\'{e}mentaires,
greff\'{e}es sur une autre surface -nanoparticule ou pointe d'un microscope de
force \cite{lee}, \cite{mazzola}, \cite{csaki}.\ Il existe une tension maximum
$F_{c}$ que peut supporter un pontage de ce genre. Dans la pr\'{e}sente note,
nous essayons de comprendre la forme de la relation $F_{c}(\ell) $, et nous
comparons les r\'{e}sultats \`{a} quelques donn\'{e}es de la r\'{e}f.
\cite{lee}.

Le mod\`{e}le utilis\'{e} est un mod\`{e}le ''d'\'{e}chelle souple''
d\'{e}crit sur la figure \ref{fig1}. Bien entendu, il faut imaginer cette
\'{e}chelle comme torsad\'{e}e pour engendrer une double h\'{e}lice, mais il
semble que l'hamiltonien de base (\'{e}q. \ref{eq1}) garde un sens pour une
forme plus r\'{e}aliste.

Le r\'{e}sultat est une distortion, qui est importante pr\`{e}s des deux
extr\'{e}mit\'{e}s de la zone appari\'{e}e, et qui s'\'{e}tend sur une
longueur $\varkappa^{-1}$ pr\`{e}s des deux extr\'{e}mit\'{e}s. On ne gagne
rien sur $F_{c}$ \`{a} cr\'{e}er des structures hybrides de longueur
$\ell>>2\varkappa^{-1}$. En comparant ces id\'{e}es aux r\'{e}sultats de la
r\'{e}f\'{e}rence \cite{lee}, on arrive \`{a} une estimation $\varkappa
^{-1}\sim6.2$ paires de bases.

\section{Introduction}

Hybridisation between short DNA sequences, grafted on two surfaces, can lead
to useful recognition systems.\ In particular, one can use an AFM tip
\cite{lee}, \cite{mazzola}, \cite{csaki}, measuring the force between tip and
substrate for a single molecule. An obvious question then arises: what is the
dependence of the rupture force $F_{c}$ on the length of the hybridized
portion? Is it useful to go to long lengths?

In the present note, we try to provide a qualitative answer to this question,
using a very crude ''ladder model'' for the hybridized portion.

\section{The ladder model}

\begin{figure}
[h]
\begin{center}
\includegraphics[
height=1.0871in,
width=4.58in
]%
{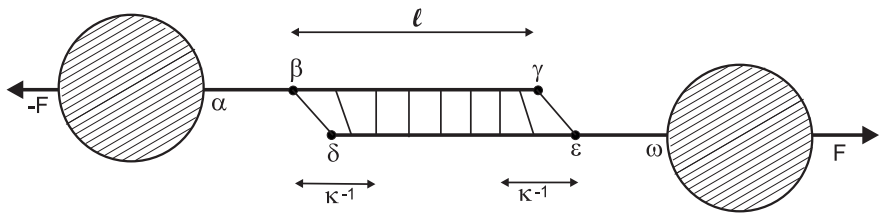}%
\caption{Two particles linked by a single bridge}%
\label{fig1}%
\end{center}
\end{figure}

We focus our attention on fig.\ref{fig1}, where the tension $F$ is fed on one
single strand at each end. A complete description of a distorted helix is, in
principle, possible but complicated.\ We shall use here the simpler ''ladder
model'' sketched on fig.\ 1. In the real world, the ladder is twisted into a
double helix. But the distribution of forces should be qualitatively the same
for both cases.%

Our starting point is a set of one dimensional displacement $u_{n}$ and
v$_{n}$ for the two sides of the ladder: ie for the members of a base pair
($n$). $u_{n}$ describes the ($\delta\epsilon$) portion (with $n$ ranging from
-$\ell/2$ to $\ell/2$) and v$_{n}$ describes the conjugate portion
($\beta\gamma$). $u_{n}$ is different from v$_{n},$ because the ladder is
distorted by the force $F$. We postulate an elastic energy:%

\begin{equation}
H=\sum_{-\ell/2}^{\infty}\frac{1}{2}Q\left(  u_{n+1}-u_{n}\right)  ^{2}%
+\sum_{-\infty}^{\ell/2}\frac{1}{2}Q\left(  \text{v}_{n+1}-\text{v}%
_{n}\right)  ^{2}+\sum_{-\ell/2}^{\ell/2}\frac{1}{2}R\left(  u_{n}%
-\text{v}_{n}\right)  ^{2} \label{eq1}%
\end{equation}

The $Q$ terms describe elongation of the backbone in one same piece:
($\alpha\gamma$) or ($\delta\omega$).\ The $R$ terms come from the coupling
between base pairs, and we expect $R$ to be weaker than $Q.$

We shall supplement the elastic description of eq. \ref{eq1} by a breaking
condition: whenever the forces inside a base pair ($n$) are larger than a
certain threshold $f_{c}$, the bond will break.\ This corresponds to:%

\begin{equation}
R\left|  \text{v}_{n}-u_{n}\right|  >f_{c} \label{eq2}%
\end{equation}

The equilibrium conditions derived from eq. (\ref{eq2}) are:%

\begin{equation}
-\frac{\partial H}{\partial\text{v}_{n}}\equiv Q\left(  \text{v}%
_{n+1}-2\text{v}_{n}+\text{v}_{n-1}\right)  +R\left(  u_{n}-\text{v}%
_{n}\right)  =0 \label{eq3}%
\end{equation}
for all indices $n$ in the interval ($-\ell/2\leqslant n\leqslant\ell
/2$).\ Outside of the interval, the $R$ term drops out.\ We shall be
concerned, in practice, with $\ell$ values significantly larger than one, and
go to the continuum limit:%

\begin{equation}
Q\frac{d^{2}\text{v}}{dn^{2}}+R\left(  u-\text{v}\right)  =(0) \label{eq4}%
\end{equation}

There are similar equations for the other sequence:%

\begin{equation}
Q\frac{d^{2}u}{dn^{2}}+R(v-u)=0 \label{eq5}%
\end{equation}

Adding (\ref{eq4}) and (\ref{eq5}), we find:%

\begin{equation}
Q\frac{d^{2}}{dn^{2}}(u+\text{v})=0 \label{eq6}%
\end{equation}

and this imposes a conservation of the total tension:%

\begin{equation}
\left.
\begin{array}
[c]{c}%
Q\frac{d}{dn}(u+\text{v})=F=constant\quad(n\leqslant\ell/2)\\
u_{n}+\text{v}_{n}=nF/Q
\end{array}
\right\}  \label{eq7}%
\end{equation}

We then turn to a discussion of the difference $\delta_{n}\equiv u_{n}-$%
v$_{n}$, which is ruled by:%

\begin{equation}
Q\frac{d^{2}\delta}{dn^{2}}-2R\delta=0\quad(\left|  n\right|  <\ell/2
\label{eq8}%
\end{equation}

The solution is a combination of exponentials: for the problem at hand, the
right combination is symmetric upon the exchange ($n\rightarrow-n$): all the
base pairs are distorted in the same direction:%

\begin{equation}
\delta_{n}=\delta_{0}\,\text{cosh}(\varkappa n) \label{eq9}%
\end{equation}%

\begin{equation}
\varkappa^{2}=2R/Q \label{eq10}%
\end{equation}

If $R<<Q$, the ''adjustment length'' $\varkappa^{-1}$ is larger than unity.

The overall solution derived from eqs (\ref{eq8}) and (\ref{eq10}) is:%

\begin{equation}
\left.
\begin{array}
[c]{c}%
u_{n}=nF/2Q+\frac{1}{2}\delta_{0}\,\text{cosh}(\varkappa n)\\
\text{v}_{n}=nF/2Q-\frac{1}{2}\delta_{0}\,\text{cosh}(\varkappa n)
\end{array}
\right\}  \label{eq11}%
\end{equation}
The relation between $\delta_{0}$ and $F$ is derived from the boundary
condition at $n=\ell/2.\;$Here, we must have:%

\[%
\begin{array}
[c]{cc}%
F & =Q(u_{n}-u_{n-1})+R(u_{n}-\text{v}_{n})
\end{array}
\]

giving:%

\begin{equation}
F=\delta_{0}\left\{  Q\varkappa\,\text{sinh}\left(  \varkappa\frac{\ell}%
{2}\right)  +2R\,\text{cosh}\left(  \varkappa\frac{\ell}{2}\right)  \right\}
\label{eq12}%
\end{equation}

The force on the last hydrogen bond ($n=\ell/2$) is $R\delta_{n}$: when we
reach the threshold $f_{1}$, this corresponds to:%

\begin{equation}
R\delta_{0}\text{cosh}(\varkappa\ell/2)=f_{1} \label{eq13}%
\end{equation}

Eq. (\ref{eq12}) then gives a global tension at threshold:%

\begin{equation}
F_{c}=2f_{1}\left\{  \varkappa^{-1}\,\text{tanh}\left(  \varkappa\frac{\ell
}{2}\right)  +1\right\}  \label{eq14}%
\end{equation}

Two limits are of interest:

\qquad a) short strands ($\varkappa\ell<1$), correspond to:%

\begin{equation}
F_{c}=f_{1}(\ell+2) \label{eq15}%
\end{equation}

$F_{c}$ increases linearly with $\ell.$

\qquad b) infinitely long strands: the rupture force reaches a maximum:%

\begin{equation}
F_{c}\rightarrow F_{m}=2f_{1}(\varkappa^{-1}+1) \label{eq16}%
\end{equation}

The force $F_{m}$ is much larger than $2f_{1}$, because a number
$\varkappa^{-1}$ of base pairs work in parallel, near each end.

Just below this maximum, we can write from eq. (\ref{eq14}):%

\begin{equation}
F_{c}/2f_{1}=\varkappa^{-1}(1-2e^{-\varkappa\ell})+1 \label{eq17}%
\end{equation}

All the base pairs in the adjustment regions (of length $\varkappa^{-1}$)
participate to the resistance, while the center portion is at rest.

Note that, when the pairing breaks at both ends ($n=\pm\ell/2$) the other ties
inside ($\left[  n\right]  <\ell/2$), also break (since $F_{c}$ is an
increasing function of $\ell$).

\section{Discussion}

\ \ \ \ \ 1) Clearly, we do not gain strength by choosing a long pairing
sequence: the optimal pairing length is 2$\varkappa^{-1}$.\ This conclusion
was achieved with a primitive ''ladder model''.\ But the existence and meaning
of the length $\varkappa^{-1}$ are probably more general.

2) The magnitude of $\varkappa^{-1}$ is unclear. If the leading coupling
between strands is due to base pairing, this should be smaller than the
covalent bonds, and $\varkappa^{-1}$ should be larger than unity.\ But we can
also think that the spring constant $R$ in eq. (\ref{eq1}) is partly controled
by the stacking of base pairs: bending the base plane by an angle $\sim
(0_{n}-$v$_{n})/D$ (where $D$ is the radius of the 2s DNA) may contribute
another energy to $R$, and reduce $\varkappa^{-1}.$

If we start from the data of ref \cite{lee}, we find that the force $F_{c}$
was roughly equal to 1.11 nanonewton for both $\ell=20$ and $\ell=16$: this
corresponds to the saturation regime ($F_{c}=F_{m}$ for $\varkappa\ell
>>1$).\ On the other hand, for $\ell=12$, $F_{c}=0.83$ nN.

Using eqs (\ref{eq16}, \ref{eq17}), this leads us to $\varkappa^{-1}=6.2$ base
pairs.\ Thus, at least for this particular example, the adjustment length
seems to be much larger than unity. The corresponding force $f_{1}$ is 17 pN.

\bigskip

\textit{Acknowledgments}: I have greatly benefited from discussions with
F.\ Brochard-Wyart,V.\ Croquette, J.\ F.\ Joanny, P.\ Pincus, E.\ Westhof and C.\ Wyart.

\end{document}